\documentclass{ceab}   
\usepackage{epsfig}     
\usepackage{graphicx}   
\usepackage{hyperref}
\usepackage{ceabbib}     
\usepackage[T1]{fontenc}
\usepackage{amsmath, amsthm, amsfonts}
\usepackage{textcomp}
\usepackage{subfig}
\usepackage{rotating}

\def\gore{CME reconstruction and interplanetary expansion}

\begin{document}

\title{3D reconstruction and interplanetary expansion of the 2010 April 3$^{rd}$ CME}

\author{M. RODARI$^{1,2}$, M. DUMBOVI\'C$^{1}$, M. TEMMER$^{1}$, L.M. HOLZKNECHT$^{1}$\\ and 
A. VERONIG$^{1}$
\vspace{2mm}\\
\it $^1$Institute of Physics, University of Graz, Austria\\ 
\it $^2$Institute of Physics G. Occhialini, University of Milano-Bicocca, Italy\\ 
 Correspondence: \href{mailto:martina.rodari@gmail.com}{martina.rodari@gmail.com},    \href{mailto:mateja.dumbovic@uni-graz.at}{mateja.dumbovic@uni-graz.at}
}

\maketitle

\begin{abstract}
We analyse the 2010 April 3$^{rd}$ CME using spacecraft coronagraphic images at different vantage points (SOHO, STEREO-A and STEREO-B).
We perform a 3D reconstruction of both the flux rope and shock using the Graduated Cylindrical Shell (GCS) model to calculate CME kinematic and morphologic parameters (\textit{e.g.} velocity, acceleration, radius).
The obtained results are fitted with empirical models describing the expansion of the CME radius in the heliosphere and compared with in situ measurements from \textit{Wind} spacecraft: the CME is found to expand linearly towards Earth. 
Finally, we relate the event with decreases in the Galactic Cosmic Ray (GCR) Flux, known as Forbush decreases (FD), detected by EPHIN instrument onboard SOHO spacecraft. We use the analytical diffusion-expansion model (ForbMod) to calculate the magnetic field power law index, obtaining a value of $\sim 1.6$, thus estimating a starting magnetic field of $\sim 0.01$ G and an axial magnetic flux of $\sim 5 \cdot 10^{20}$ Mx at 15.6 R$_\odot$ .

\end{abstract}


\section{Introduction}
Coronal Mass Ejections (CMEs) are magnetic structures which erupt in the Sun's atmosphere, propagate into the interplanetary space and can have significant impact on the Earth's magnetosphere. Studying the initial and evolutionary properties of CMEs improves our efforts in space weather forecasting.

Previous studies already analyzed the 2010 April $3^{rd}$ CME, focusing on the interplanetary propagation \citep{moestl2010} or on the 3D CME reconstruction \citep{wood2011}.
We combine the 3D CME reconstruction using the Graduated Cylindrical Shell (GCS) model \citep{Thernisien2011} with the interplanetary propagation given by the Drag-Based Model \citep{vrsnak2013} in the ensemble mode \citep[DBEM;][]{dumbovic2018_DBEM})\footnote{available under \url{http://swe.ssa.esa.int/heliospheric-weather}} to associate this CME with an ICME observed in situ on 2010 April 5$^{th}$. We note that our association is in agreement with \cite{moestl2010}.

Finally, we analyse the CME expansion using \textit{in-situ} plasma and interplanetary magnetic field measurements by \textit{Solar Wind Experiment} \citep{ogilvie95} and \textit{Magnetic Field Experiment} \citep{lepping95}, respectively. In addition, we use galactic cosmic rays (GCRs) measurements by F-detector of the SOHO-EPHIN instrument \citep{muller-mellin95} as indirect CME probes through their interaction with the flux rope magnetic structure.

\section{Data and Methods}

The 2010 April 3$^{rd}$ CME originated from the NOAA active region AR11059 located at S25W03 and was associated with a B7.4 flare starting at 09:14 UT.
\begin{figure}[hbtp]
  \begin{center}
   \includegraphics[scale=0.3]{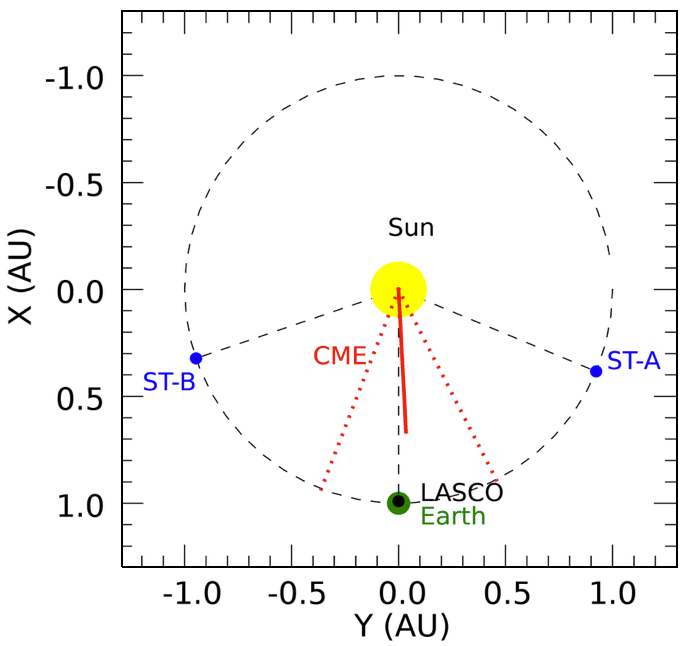}%
  \end{center}
  \caption{CME direction (Stonyhurst long=3\textdegree, lat=$-$28\textdegree) with respect to Earth, STERO-A and -B (STEREO-A and -B separation angle is 139\textdegree).}\label{fig:cmeprop}
\end{figure} 

\begin{figure}[htb]
	\includegraphics[width=\columnwidth]{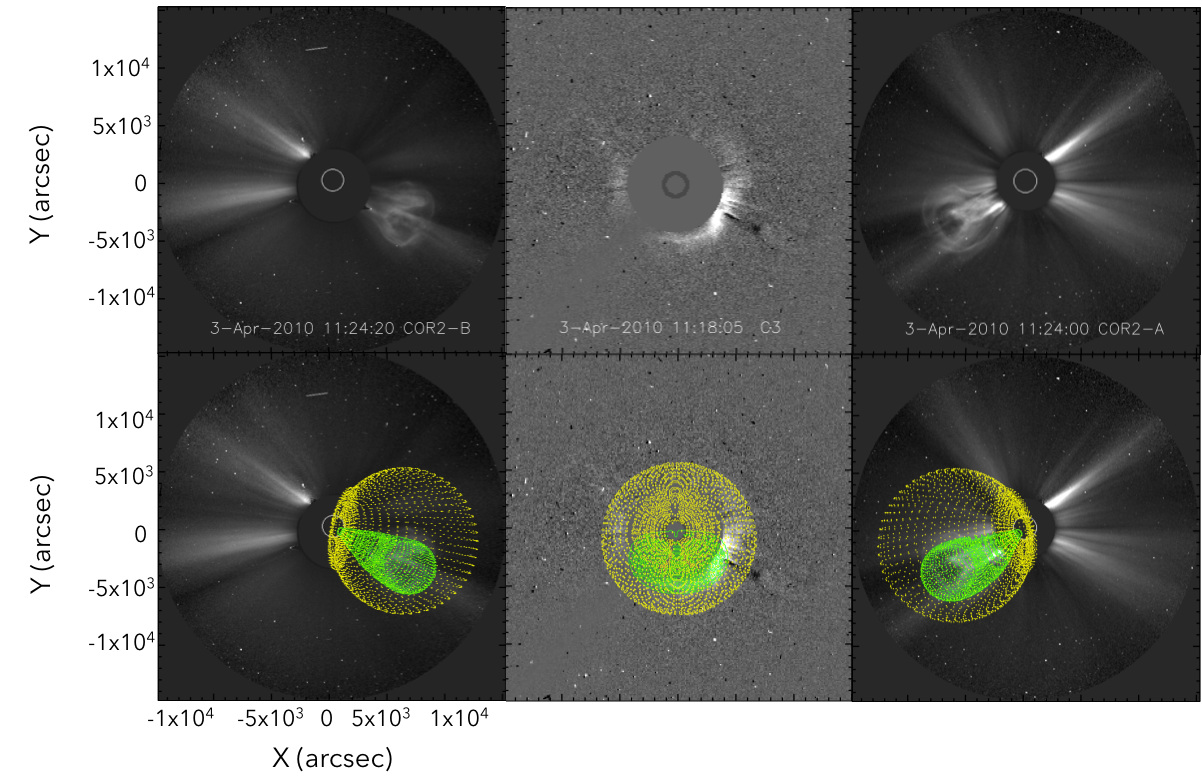}%
	\caption{GCS reconstruction of the CME at 11:24 UT using STEREO-B (left), LASCO (middle), and STEREO-A (right). We track the outer bright rim visible in STEREO as the flux rope (FR, green structure) and defined the almost-spherical structure visible in LASCO  as the shock (yellow).}\label{fig:gcs}
\end{figure}

We use white light coronagraphic images from the \textit{Sun Earth Connection Coronal and Heliospheric Investigation} (SECCHI) 
instrument suite \citep{secchi} onboard the \textit{Solar TErrestrial RElations Observatory} (STEREO) mission and \textit{Large Angle Spectrometer Coronagraph} \citep[LASCO;][]{lasco} onboard \textit{SOlar and Heliospheric Observatory} (SOHO). 
The CME geometry is obtained using the graduated cylindrical shell (GCS) reconstruction model \citep{Thernisien2006,thernisien2009,Thernisien2011}, which represents the CME as a "hollow-croissant" flux rope (see green structure in \figurename~\ref{fig:gcs}).
Since we are considering a \textit{self-similar expansion}, we only increase the height in time and keep all other parameters fixed (the half angle is 24\textdegree, the aspect ratio is 0.29 and the tilt angle is 1.7\textdegree).

\begin{figure}[hbt]
	\includegraphics[width=\columnwidth]{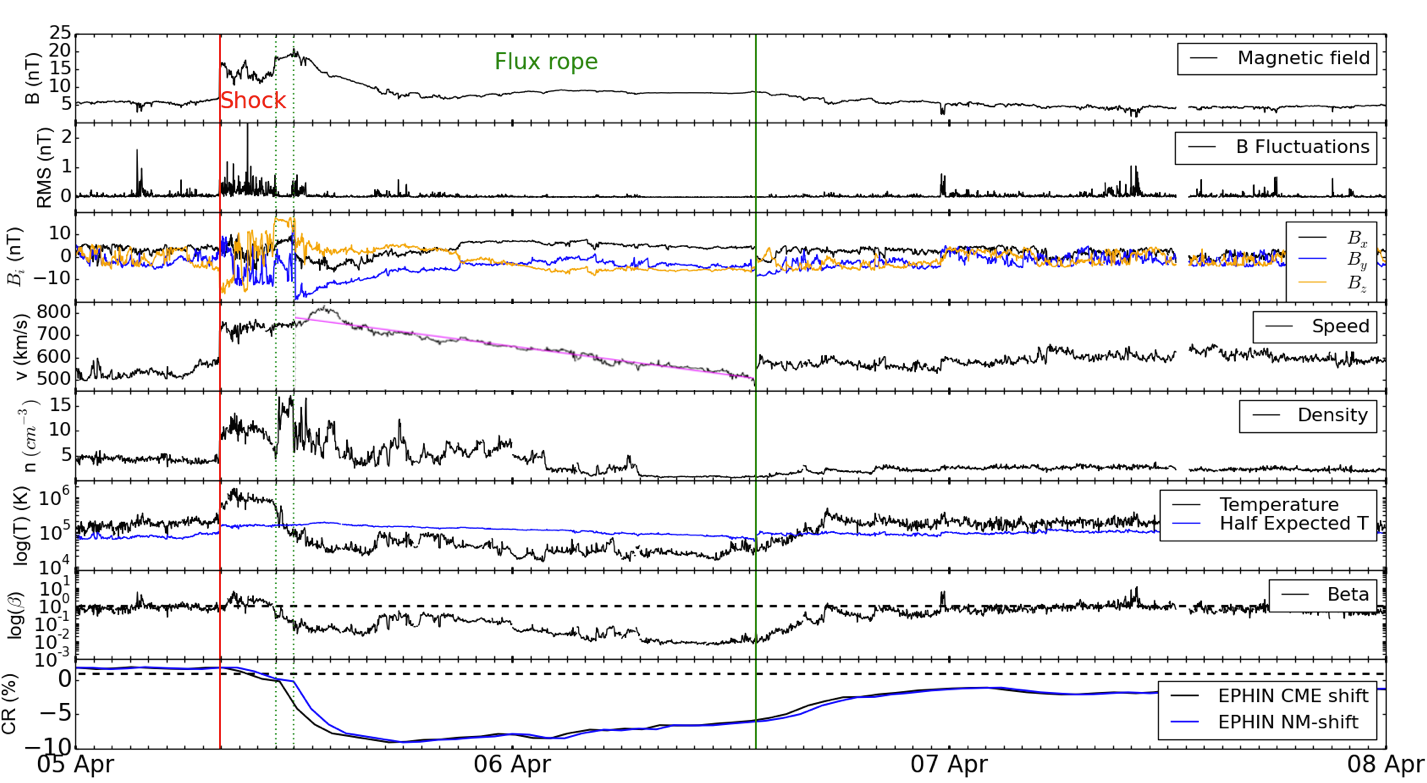}%
	\caption{Top to bottom: \textit{in situ} measurements of the IMF strength, fluctuations, and IMF components, plasma speed, density, temperature and plasma beta. The bottom-most panel represents the decrease in GCR count as measured by SOHO-EPHIN F-detector (data are shifted to Earth using the ICME speed (in black) and the SOPO neutron monitor (NM) data (in blue)). The red line marks the shock, whereas the green lines mark the beginning/end of the FR. The dotted green lines define the uncertainty range of the FR arrival.}\label{fig:wind1}
\end{figure}

We then analyse the solar wind (SW) and interplanetary magnetic field (IMF) data of the associated ICME detected by the \textit{Wind} spacecraft. The shock arrived on April the 5$^{th}$, indicated by a sudden increase in the IMF intensity, SW temperature and density (see red line in \figurename~\ref{fig:wind1}), followed by a sheath region identified through the strong IMF fluctuations \citep[see \textit{e.g.}][and references therein]{kilpua2017}.
Four hours later magnetic cloud signatures are detected which last almost one day: a linear decrease of the plasma speed (indicating expansion), temperature below half the expected and a low plasma beta parameter \citep{wang05,kilpua2017}. However, there is no clear rotation in the magnetic field components, possibly related to the spacecraft trajectory through the leg of the FR \citep{moestl2010,wood2011}.

We analyse the CME expansion, assuming that it follows a power-law relation, where the final radius is a scaled version of the initial one \citep[see Eq.13 in][]{dumb2018_F}. We use \textit{in situ} measurements to calculate the power-law index $n_a$, using two different methods (see first row of Table 1). In method 1 we use the empirical relation found in \cite{guli2012}, which connects the difference between the front and rear FR speed and its transit time.
In method 2 we use the observed linear fit slope of the SW speed \citep{demdasso2009}. Using the power-law relation we back-extrapolate the FR radius derived by the two methods to obtain the "initial" FR radius at the distance where GCS reconstruction was performed and compare to the GCS-calculated FR radius (see 2$^{nd}$ row of Table 1).

Considering a power-law decrease of the magnetic field inside the flux rope (see eq.14 in \cite{dumb2018_F}), we calculate the power-law index $n_B$, using the diffusion-expansion Forbush decrease (FD) model \textit{ForbMod} \citep{dumb2018_F}. With \textit{ForbMod} we can use the FD amplitude measured by the SOHO/EPHIN F-detector $|FD|_{max}$ and $n_a$ obtained from the \textit{in situ} measurements to estimate $n_B$:
\begin{equation}
 n_{B} = 2n_{a}-1-\frac{\alpha_{1}^{2}}{ln |FD|_{max}} \frac{D(t)}{a_{0}^{2}} \left(  \frac{R(t)}{R_{0}}\right)^{-2n_{a}} t
\end{equation}
\noindent where $\alpha_1$ is first positive root of the order zero Bessel function, $a_0$ is the initial CME radius, $R_0$ is the initial CME height, $R(t)$ is the traveled distance, $t$ is the transit time and $D(t)$ is the diffusion coefficient which is also given by the power-law behavior, scaled to the radial perpendicular diffusion coefficient at Earth given in \cite{potg2013}.

\section{Results and discussion}
The final GCS reconstruction is performed at $15.6 \pm 0.8$ R$_\odot$ where we already observe CME deceleration obtaining a  speed of $920\pm 200$ km/s and a radius of $3.1 \pm 0.2$ R$_\odot$. The power-law indices $n_a$ obtained from two different methods are very similar, close to one (row 1 in Table 1) yielding the initial FR radius close to the GCS-reconstructed value (row 2 in Table 1). The errors are obtained taking the minimum and maximum value of the parameters and calculating the expected value range. 

\begin{table}[hbt]
\centering
	\begin{tabular}{|c|c|c|}
	\hline
	 & \textit{1st method} &\textit{2nd method} \\
     \hline
     \textbf{$n_a$} & $0.98 \pm 0.03$ & $0.99 \pm 0.02$ \\
     \textbf{$a_0$}& $(3.4 \pm 0.5) R_\odot$ & $(3.3 \pm 0.4) R_\odot$ \\
     \textbf{$n_B$} & $1.6 \pm 0.3$ & $1.6 \pm 0.2$ \\
     \textbf{$x =n_B-2n_a$}   & $-0.4$ & $-0.4$ \\
     \textbf{$B_0$} & $(0.01 \pm 0.007)$ G & $(0.009 \pm 0.005)$ G \\
     \textbf{$\Phi_{ax}=1.4~B_0~r_0^2$} & $\sim 4.98\cdot 10^{20}$ Mx & $\sim 4.86 \cdot 10^{20}$ Mx \\
     \hline
     \textbf{$B_0$} & \multicolumn{2}{|c|}{0.010 - 0.018 (0.013) G}\\
     \textbf{$a_0$} & \multicolumn{2}{|c|}{0.5 - 3.1 (1.2) R$_\odot$} \\
     \hline
\end{tabular}
\caption{Calculated values using methods 1 and 2 (top to bottom): the FR expansion power-law index, the power-law back-extrapolated initial FR radius, the magnetic field power-law index, the expansion type, the initial FR central magnetic field, the axial magnetic flux, and the initial FR central magnetic field and radius calculated from the empirical relations in the inner heliosphere obtained from \cite{leitner2007}.}
\label{tab:results}
\end{table}

Both methods give same results for $n_B$, as can be seen in row 3 of Table 1 and we note that this value is in agreement with statistical studies \citep{leitner2007,guli2012}. Moreover we calculate the expansion type, $x$ (see row 4 of Table 1) where a negative value indicates an effective increase of the axial magnetic flux inside the CME. This is related to the fact that the expansion of the FR radius is very quick compared to the drop in the central magnetic field \citep[for a more detailed discussion on the expansion types we refer the reader to][]{dumb2018_F}. Using the \textit{in situ} IMF measurements we back-extrapolate the magnetic field using the power-law index $n_B$ to estimate the initial FR central magnetic field (row 5 of Table 1) and axial magnetic flux, $\Phi_{ax}$ (row 6 in Table 1). We note that the values obtained for the $\Phi_{ax}$ are found around typically expected values reported in  previous studies \citep[\textit{e.g.}][]{devore2000,temmer2017}. Finally, we use the empirical power-law equations by \cite{leitner2007} to calculate the initial FR radius and magnetic field using \textit{in situ} measurements and we find good agreement with our calculated values.

\section{Conclusion and summary}
In this work we estimate the initial magnetic field and axial flux inside the CME using a set of multi-spacecraft and multi-instrument observations, as well as modelling.

Using 3D GCS reconstruction from white light coronagraphs and \textit{in-situ} measurements we constrain the power-law index of the FR size with two different methods which yield similar results close to one. Furthermore, using GCR measurements and FD model \textit{ForbMod} we obtain the power-law index of the FR central magnetic field to be 1.6 which allows us to estimate the initial central FR field and magnetic flux.

We find that this particular CME shows expansion parameters which are typically found in statistical studies, whereas their combination, \textit{i.e.} expansion type, shows that the FR size is expanding much faster than the decrease rate of the magnetic field. This indicates that the magnetic flux inside the flux rope is effectively increasing as it moves away from the Sun. The method is found to provide reasonable results comparable to other studies. A future statistical study is planned to better define the validity of this approach.

\section*{Acknowledgements} 

This research was possible with the support of the Erasmus+ for Traineeship program of the European Union.
M.D. acknowledges funding from the EU H2020 MSCA grant agreement No 745782 (project ForbMod). We acknowledge the use of the SOHO/EPHIN data, supported under Grant 50 OC 1302 by the German Ministry of Science.

\bibliographystyle{ceab}
\bibliography{sample}

\end{document}